\documentclass[preprint]{aastex701}

\begin{document}

\title{Palomar and Apache Point Spectrophotometry of Interstellar Comet 3I/Atlas}

\author[0000-0003-4778-6170]{Matthew Belyakov}
\affiliation{Division of Geological and Planetary Sciences, California Institute of Technology, Pasadena, CA 91125, USA}
\email{mattbel@caltech.edu}
\correspondingauthor{Matthew Belyakov}

\author[0000-0002-4223-103X]{Christoffer Fremling}
\affiliation{Division of Physics, Math, and Astronomy, California Institute of Technology, Pasadena, CA 91125, USA}
\email{fremling@caltech.edu}

\author[0000-0002-3168-0139]{Matthew J. Graham}
\affiliation{Division of Physics, Math, and Astronomy, California Institute of Technology, Pasadena, CA 91125, USA}
\email{mjg@caltech.edu}

\author[0000-0002-4950-6323]{Bryce T. Bolin}
\affiliation{Eureka Scientific, Oakland, CA, 94602}
\email{bbolin@eurekasci.com}

\author[0000-0001-6098-2235]{Mukremin Kilic}
\affiliation{University of Oklahoma Department of Physics and Astronomy, 440 W. Brooks St.
Norman, OK 73019}
\email{mukreminkilic@gmail.com}

\author[0009-0009-9105-7865]{Gracyn Jewett}
\affiliation{University of Oklahoma Department of Physics and Astronomy, 440 W. Brooks St.
Norman, OK 73019}
\email{gjewett@ou.edu}

\author[0000-0002-9548-1526]{Carey M. Lisse}
\affiliation{Johns Hopkins University Applied Physics Laboratory, 11100 Johns Hopkins Rd, Laurel, MD 20723, USA}
\email{carey.lisse@jhuapl.edu}

\author[0000-0002-7053-5495]{Carl Ingebretsen}
\affiliation{Department of Physics and Astronomy, Johns Hopkins University, Baltimore, MD 21218, USA}
\email{cingebr1@jh.edu}

\author[0000-0002-7451-4704]{M. Ryleigh Davis}
\affiliation{Division of Geological and Planetary Sciences, California Institute of Technology, Pasadena, CA 91125, USA}
\email{rdavis@caltech.edu}

\author[0000-0001-9665-8429]{Ian Wong}
\affiliation{NASA Goddard Space Flight Center, 8800 Greenbelt Road, Greenbelt, MD 20771, USA}
\email{iwong@mit.edu}

\begin{abstract}
On July 1st 2025 the third interstellar object, 3I/ATLAS or C/2025 N1 (ATLAS), was discovered, with an eccentricity of $e=6.15 \pm 0.01$ and perihelion of $q=1.357\pm0.001$ au. We report our initial visible to near-infrared (420-1000 nm) spectrophotometry of 3I/ATLAS using both the Palomar 200 inch telescope and Apache Point Observatory. We measure 3I/Atlas to have a red spectral slope of 19\%/100 nm in the 420-700 nm range, and a more neutral 6\%/100 nm slope over 700-1000 nm. We detect no notable emission features such as from C$_2$. 
\end{abstract}

\section{Introduction} 
Interstellar objects (ISOs) are macroscopic km-scale bodies passing through the Solar System on orbits inconsistent with an Oort Cloud or Kuiper Belt source due to their significant hyperbolic eccentricities. Suggested to be the scattered remnants of planetary formation around other stars, ISOs provide a unique, albeit transient, bridge between observations of aggregate dust in debris disks and individual solar system asteroids and comets \citep{Jewitt2023ARA&A}. 

The two previous interstellar objects discovered, 1I/'Oumuamua and 2I/Borisov, gave distinct previews of the physical properties and composition of the ISO population. 1I/'Oumuamua displayed no detectable activity yet had a notable non-gravitational acceleration whose cause has still not been definitively established \citep{Meech2017Natur, Micheli2018Natur,Seligman2020ApJL, Desch2021JGRE}. By contrast, 2I/Borisov behaved more similarly to typical Oort cloud comets, showing signs of clear activity as it passed through the Solar system \citep{Jewitt2019ApJL, deLeon2020MNRAS, Guzik2020NatAs}. 

3I/Atlas, discovered on July 1st using the ATLAS network of telescopes, is unambiguously an interstellar object owing to its significant eccentricity ($e\simeq$ 6). This new interloper shows clear activity with an extended surface brightness profile \citep{Bolin20253I, Seligman2025arXiv}, promising future detection of cometary volatiles as it approaches perihelion, though none have been reported at the time of submission of this research note \citep{Opitom2025arXiv, Seligman2025arXiv}. To add to the growing accumulation of information of this new ISO, we present preliminary spectrophotometric characterization of 3I/Atlas using the Palomar 200 inch telescope (P200) and the Apache Point Observatory (APO).

\begin{figure*}[h]
    \centering
    \includegraphics[width=1\linewidth]{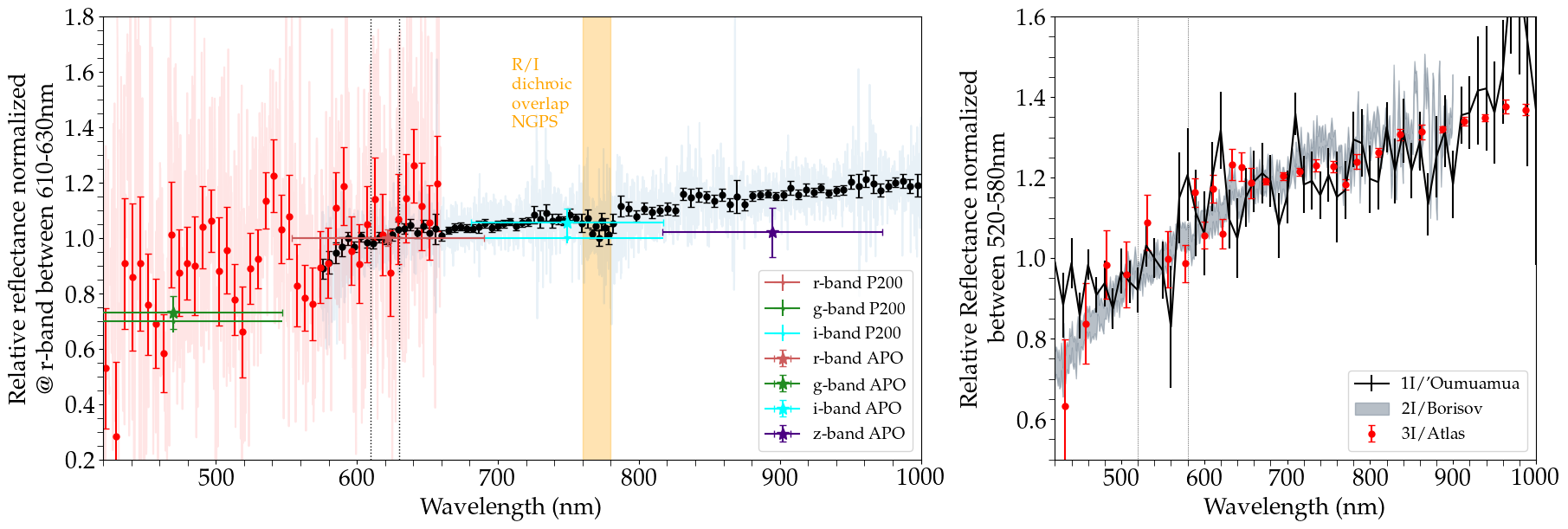}
    \caption{Left: Palomar and Apache Point imaging and spectroscopy. 3I/Atlas has a red spectrum in the visible that becomes more neutral at longer wavelengths. Right: Comparison between our combined and binned 3I/Atlas spectrum and 1I from \cite{Fitzsimmons2018NatAs} and 2I from \cite{deLeon2020MNRAS}. 3I/Atlas appears roughly consistent with 2I/Borisov. Area used to normalize spectra shown as faint dotted lines. All error bars are 1-sigma.}
    \label{fig:plot}
\end{figure*}

\section{Observations and Data Reduction}
We observed 3I/Atlas on 3 July 2025 UTC with the P200 using the Next Generation Palomar Spectrograph \citep[NGPS\footnote{\url{https://www.astro.caltech.edu/ngps}},][]{Jiang2018SPIE}. NGPS currently operates in simultaneous two-channel mode covering $R$ band (580-780 nm) and $I$ band (760-1040 nm). The spectrograph has an image slicer which captures light outside the center slit onto two secondary slits. At the 1.5$\arcsec$ slit width used for our observations, 80\% of the light enters the center slit, and 20\% into the auxiliary slits. The instrument also has a $4'\times4'$ guider camera suitable for broadband photometry with $g$, $r$, and $i$ filters at a 0.26$\arcsec$ pixel scale. The seeing at the time of observations was between 1.6-1.8$\arcsec$, measured by determining the FWHM of the spectral trace in the spatial direction, and was identical to the image-measured seeing. We also obtained spectra of the G2V solar analog HIP088235 which we used to correct the comet's spectrum for telluric features and slope.

To reduce the NGPS spectra, we first bias-subtracted and flat-fielded the 2D spectra and determined the comet's trace. We then extracted along each trace using a 12 pixel width, subtracting at each wavelength the background defined as the median of 10-20 pixels on either side of the trace. We then combined the signal from the center and bottom slits -- the top slit did not show the comet. We combine the $R$ and $I$ spectra by normalizing at their shared wavelengths. This process was repeated for our solar analog.

We observed 3I/Atlas from APO on 6 Aug 2025 UTC, obtaining $g/r/i/z$-band images with the ARCTIC camera \citep{Huehnerhoff2016SPIE}, with a measured $\sim$1.5\arcsec\ seeing in r-band. We also obtained a spectrum using the KOSMOS spectrometer \citep{Kadlec2024} to complement our Palomar observations. KOSMOS was used with the blue grism, center slit position, and 2.1$\arcsec$ slit covering the 380 nm - 660 nm spectral range with a spectral resolution of $\sim$500. The KOSMOS data were detrended using biases and arc lamps. We similarly observed the solar analog HIP088235 for telluric feature and slope correction. 

The imaging results from both Palomar and APO are given in \citep{Bolin20253I}. The raw and reduced imaging and spectroscopy data have been made public at \cite{belyakov_fremling_graham_mukremin_bolin_jewett_lisse_2025}.

\section{Results}
We show the combined photometric and spectroscopic data in the left panel of \autoref{fig:plot}. Our photometry and spectra are roughly consistent, finding a strongly red slope between 420-700 nm. We measure this slope to be 18.9$\pm$0.1 \%/100 nm based on the covariance matrix of the linear fit to the data over the wavelength range. By contrast, we find a 6.2$\pm$0.1 \%/100 nm slope across 700-1000 nm. We also find an extension of the object beyond stellar point source out to 5$\arcsec$ \citep{Bolin20253I}, indicating activity. We do not see any obvious emission from C$_2$ or CO$^+$, typical gas species found in cometary comae \citep{Schleicher2004come}. We will provide upper limit estimates of gas production in a forthcoming article.

The lack of emission from common volatiles is not entirely unexpected as we observed 3I/Atlas at a heliocentric distance of 4.3 au from the sun, outside of where water ice begins to rapidly sublimate. By contrast, 2I/Borisov showed evidence of distant activity beyond the water-ice sublimation line \citep[][]{Bolin20202I}, significant seasonal variations in its activity \cite[][]{Bolin2020HST}, and a high carbon dioxide abundance in its coma \citep[][]{Cordiner2020}. First measurements of 3I/Atlas indicate activity due to its appearance as an extended point source, however the cause of activity remains unclear. Future observations of 3I taken by both ground and space-based facilities prior to and after its perihelion approach are required to understand any activity drivers and 3I's evolutionary state.

\bibliography{sample701}{}
\bibliographystyle{aasjournal}
\end{document}